\documentclass{article}

\usepackage{arxiv}

\usepackage[utf8]{inputenc} 
\usepackage[T1]{fontenc}    
\usepackage{hyperref}       
\usepackage{url}            
\usepackage{booktabs}       
\usepackage{amsfonts}       
\usepackage{nicefrac}       
\usepackage{microtype}      
\usepackage{lipsum}		
\usepackage{graphicx}
\usepackage{doi}
\usepackage{amsmath,amssymb,amsfonts}
\usepackage{algorithmic}
\usepackage{graphicx}
\usepackage{algorithm,algorithmic}
\usepackage{hyperref}
\usepackage{tabularx}  
\usepackage{booktabs} 
\hypersetup{hidelinks=true}
\usepackage{textcomp}
\usepackage{amsmath} 
\DeclareMathOperator*{\argmin}{arg\,min}  
\DeclareMathOperator{\Exp}{Exp}
\DeclareMathOperator{\Log}{Log}

\title{Riemannian Geometry-Based EEG Approaches: A Literature Review}

\author{
    \href{https://orcid.org/0009-0004-4729-7128}{\includegraphics[scale=0.1]{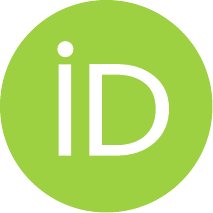}\hspace{1mm}Imad Eddine Tibermacine}\textsuperscript{1} \\
    \and
    \href{https://orcid.org/0000-0002-1846-9996}{\includegraphics[scale=0.1]{orcid.pdf}\hspace{1mm}Samuele Russo}\textsuperscript{2} \\
    \and
    \href{https://orcid.org/0000-0001-7021-1194}{\includegraphics[scale=0.1]{orcid.pdf}\hspace{1mm}Ahmed Tibermacine}\textsuperscript{3} \and
    \href{https://orcid.org/0000-0001-8684-4754}{\includegraphics[scale=0.1]{orcid.pdf}\hspace{1mm}Abdelaziz Rabehi}\textsuperscript{4} \and
    \href{https://orcid.org/0000-0002-2542-8391}{\includegraphics[scale=0.1]{orcid.pdf}\hspace{1mm}Bachir Nail}\textsuperscript{4} \and
    \href{https://orcid.org/0009-0002-9687-4943}{\includegraphics[scale=0.1]{orcid.pdf}\hspace{1mm}Kamel Kadri}\textsuperscript{5} \and
    \href{https://orcid.org/0000-0002-3336-5853}{\includegraphics[scale=0.1]{orcid.pdf}\hspace{1mm}Christian Napoli}\textsuperscript{1,6,7}
}

\date{}

\hypersetup{
pdftitle={A template for the arxiv style},
pdfsubject={q-bio.NC, q-bio.QM},
pdfauthor={David S.~Hippocampus, Elias D.~Striatum},
pdfkeywords={First keyword, Second keyword, More},
}

\begin{document}
\maketitle

\maketitle

\noindent \textsuperscript{1}Department of Computer, Automation and Management Engineering, Sapienza University of Rome, Rome, Italy \{\texttt{tibermacine@diag.uniroma1.it}, \texttt{cnapoli@diag.uniroma1.it}\}\\
\noindent \textsuperscript{2}Department of Psychology, Sapienza University of Rome, Rome, Italy \{\texttt{samuele.russo@uniroma1.it}\}\\
\noindent \textsuperscript{3}Department of Computer Science, University of Biskra, Biskra, Algeria \{\texttt{ahmed.tibermacine@univ-biskra.dz}\}\\
\noindent \textsuperscript{4}Faculty of Science and Technology, University of Djelfa, Djelfa, Algeria \{\texttt{abdelaziz.rabehi@univ-djelfa.dz}, \texttt{bachir.nail@univ-djelfa.dz}\}\\
\noindent \textsuperscript{5}Polytechnic Military School, Algiers, Algeria \{\texttt{dr.kadri.kamel@gmail.com }\}\\
\noindent \textsuperscript{6}Institute for Systems Analysis and Computer Science, Italian National Research Council, Italy\\
\noindent \textsuperscript{7}Department of Computational Intelligence, Czestochowa University of Technology, Poland \\

\begin{abstract}
The application of Riemannian geometry in the decoding of brain-computer interfaces (BCIs) has swiftly garnered attention because of its straightforwardness, precision, and resilience, along with its aptitude for transfer learning, which has been demonstrated through significant achievements in global BCI competitions. This paper presents a comprehensive review of recent advancements in the integration of deep learning with Riemannian geometry to enhance EEG signal decoding in BCIs. Our review updates the findings since the last major review in 2017, comparing modern approaches that utilize deep learning to improve the handling of non-Euclidean data structures inherent in EEG signals. We discuss how these approaches not only tackle the traditional challenges of noise sensitivity, non-stationarity, and lengthy calibration times but also introduce novel classification frameworks and signal processing techniques to reduce these limitations significantly. Furthermore, we identify current shortcomings and propose future research directions in manifold learning and riemannian-based classification, focusing on practical implementations and theoretical expansions, such as feature tracking on manifolds, multitask learning, feature extraction, and transfer learning. This review aims to bridge the gap between theoretical research and practical, real-world applications, making sophisticated mathematical approaches accessible and actionable for BCI enhancements.
\end{abstract}


\section{Introduction}
Brain-computer interfaces \cite{a9} revolutionize the field of human-machine interactions by enabling direct communication pathways between the brain and external devices, bypassing traditional muscular outputs\cite{a1}. Predominantly, BCIs utilize electroencephalography (EEG) due to its cost-effectiveness and minimal computational demands\cite{a2}. EEG-based BCIs convert brain activity, measured through EEG signals, into computer commands, leveraging classification algorithms like the common spatial pattern (CSP) to recognize signal patterns associated with specific brain activities such as motor imagery\cite{a3}. Despite their promise, EEG-based BCIs face critical challenges including high signal variability, noise, non-stationarity, and the high dimensionality of data spaces\cite{a4}.

The domain of BCI has expanded its utility beyond basic research, offering profound applications in rehabilitation, assistive technologies, and adaptive human-computer interfaces that respond to the user's mental states\cite{a5}. Yet, despite these advancements, BCIs are predominantly still in the prototype stage, rarely deployed outside laboratory settings due to their low robustness and the extensive calibration required for each user\cite{a6}. Current BCIs suffer from poor reliability, often misinterpreting user commands, which significantly reduces accuracy and information transfer rates even under controlled conditions\cite{a7}. This sensitivity to environmental noise and the inherent non-stationarity of EEG signals further complicates their practical application\cite{a8}.

In the rapidly evolving field of BCIs, deep learning has emerged as a transformative force, enabling sophisticated methods for classifying EEG data\cite{a10}. These advancements hold the promise of revolutionizing the interaction between humans and machines by interpreting neural activities directly through EEG signals\cite{a11}. Despite these advances, the development of BCIs faces significant challenges, notably the limited availability of large, annotated training datasets, which are crucial for training robust deep learning models\cite{a12}. In response, researchers have increasingly turned to generative modeling, a burgeoning area within machine learning, to augment existing datasets through the synthesis of synthetic EEG data\cite{a13}. This approach not only enhances the diversity of training examples but also addresses the gap caused by scarce data resources\cite{a14}.

Achieving practical usability of BCIs requires that they be robust across different contexts, users, and over time, all while maintaining minimal calibration requirements\cite{a15}\cite{a24}. Addressing these multifaceted challenges necessitates a comprehensive strategy: identifying new, reliable neurophysiological markers at the neuroscience level\cite{a16}; improving user training to stabilize EEG pattern generation at the human interaction level\cite{a17}; and advancing signal processing techniques to enhance feature extraction\cite{a18} and classifier robustness at the computational level\cite{a19}\cite{a25}. Prominently, the application of Riemannian geometry\cite{a20} to handle EEG covariance matrices has shown considerable promise\cite{a21}. This geometric approach not only accommodates the non-Euclidean characteristics of EEG data but also leverages the intrinsic geometric properties of these data to enhance classification accuracy\cite{a22}\cite{a23}.

Traditional approaches like Fourier analysis\cite{a26} frequently fail due to the inherently nonstationary nature of EEG signals. This limitation has paved the way for the adoption of time-frequency analysis techniques, such as the wavelet transform, which are now being integrated with spatial filtering\cite{a27} methods like the CSP to improve the localizability of rhythmic EEG components. Such advancements are critical for enhancing the efficacy of motor imagery (MI) EEG classifiers, which are vital for real-time BCI applications\cite{a28}.

Moreover, a significant aspect of reducing signal variability and enhancing classifier performance in BCIs\cite{a29} involves focusing on covariance matrices, typically handled as elements within Euclidean spaces. However, considering these matrices in their naturally occurring Riemannian manifold\cite{a30}, which better reflects their symmetric and positive definite (SPD) properties\cite{a31}, can substantially optimize the computational processes. Techniques such as Principal Components Analysis (PCA)\cite{a32} and Canonical Correlation Analysis (CCA)\cite{a33} have traditionally exploited covariance estimates for spatial filtering in BCIs\cite{a34}. By refining how these matrices are treated—acknowledging their Riemannian structure—we can significantly enhance the spatial and temporal resolution of EEG signal analysis\cite{a35}.

This paper presents a detailed review focused on the applications of Riemannian Geometry and deep learning-based Riemannian geometry in the domains of BCIs and EEG. We explore how these advanced methodologies enhance BCI design by addressing core challenges inherent in traditional systems. Our discussion extends to the integration of geometric deep learning with generative modeling and sophisticated signal processing techniques. We propose new research directions and outline strategic advancements necessary for developing BCIs that are not only efficacious in controlled settings but also robust and accessible for practical, everyday applications.

\section{Brain-Computer Interface}

Brain-computer interfaces that utilize oscillatory neural activity are a significant branch of non-invasive systems for neurocommunication and control\cite{a36}. These interfaces typically leverage the modulation of brain rhythms captured via EEG, and are commonly employed in motor imagery-based BCIs\cite{a37}. Such BCIs extensively use the CSP algorithm combined with Linear Discriminant Analysis  for signal classification, making them highly effective for decoding user intentions based on neural activity patterns\cite{a38}.

The CSP algorithm is pivotal for maximizing the variance of signals from one mental state while simultaneously minimizing it for another\cite{a39}. This is achieved by designing spatial filters that enhance the discriminability between two contrasting classes of brain activity. Mathematically, the optimization of CSP filters, represented as column vectors \(\mathbf{w}\), is framed as an eigenvalue problem aimed at maximizing the following objective function:
\begin{equation}
    J_{\text{CSP}}(\mathbf{w}) = \frac{\mathbf{w}^T C_1 \mathbf{w}}{\mathbf{w}^T C_2 \mathbf{w}}
\end{equation}
Here, \(C_1\) and \(C_2\) are the averaged covariance matrices of the EEG signals filtered for two different classes, which are computed as:
\begin{equation}
    C_j = \frac{1}{N_j} \sum_{i=1}^{N_j} \mathbf{S}_j^i = \frac{1}{N_j} \sum_{i=1}^{N_j} \mathbf{Z}_j^i (\mathbf{Z}_j^i)^T
\end{equation}
where \(N_j\) is the number of trials for class \(j\), and \(\mathbf{Z}_j^i\) represents the \(i\)-th trial from class \(j\) matrix in a channel-filtered EEG dataset\cite{a40}.

The CSP algorithm essentially reduces the multi-channel EEG data into a lower-dimensional space\cite{a41}, where the spatial filters \(\mathbf{w}\) corresponding to the largest (or smallest) eigenvalues capture the most (or least) discriminative features of the EEG signals. 

Upon obtaining the optimal spatial filters, the transformed feature \(f_i\) for each trial is calculated using:
\begin{equation}
    f_i = \log(\mathbf{w}^T S_i \mathbf{w})
\end{equation}
where \(S_i\) is the covariance matrix of the current trial. These logarithmically transformed variances are then fed into an LDA classifier, which constructs a hyperplane to distinguish between the classes by solving:
\begin{equation}
    a = \mathbf{C}^{-1}(\mu_1 - \mu_2), \quad b = -\frac{1}{2} (\mu_1 + \mu_2)^T a
\end{equation}
where \(\mu_1\) and \(\mu_2\) are the mean feature vectors of each class, and \(\mathbf{C}\) is the pooled covariance matrix of the features. LDA assumes Gaussian class conditional densities with equal covariance matrices for both classes\cite{a42}, which simplifies the computation of the decision boundaries\cite{a43}\cite{a44}.

It is crucial to note that while CSP and LDA provide a robust framework for feature extraction and classification\cite{a45}, their efficacy depends on accurately estimated covariance matrices, which require sufficient trial data to avoid overfitting. Moreover, the assumption of linear separability\cite{a46} by LDA may not always hold in complex brain activity patterns, suggesting the need for more sophisticated nonlinear classifiers or deep learning approaches in more advanced BCI systems\cite{a47}\cite{a48}.

The mathematical rigor involved in optimizing CSP and employing LDA classifiers in oscillatory activity-based BCIs\cite{a49} showcases the intricate balance between statistical assumptions and practical EEG data characteristics\cite{a50}. Enhancements in computational strategies and model robustness are continually evolving, aiming to improve the adaptability and accuracy of BCIs in real-world applications\cite{a51}.

\section{Riemannian Geometry Concepts}

In this section, we introduce the mathematical notations and basic definitions fundamental to our study on EEG preprocessing. Scalars are denoted by lowercase letters (e.g., \(a\)), vectors by boldface lowercase (e.g., \(\mathbf{v}\)), representing column vectors unless stated otherwise, matrices by boldface uppercase (e.g., \(\mathbf{M}\)), and tensors of order three or higher by Euler script letters (e.g., \(\mathcal{T}\)). The space \(\mathbb{R}^n\) represents the \(n\)-dimensional real vector space, and \(\mathbb{R}^{n\times n}\) denotes all \(n \times n\) real matrices. Vector \(\mathbf{e}_n\), comprising all ones, belongs to \(\mathbb{R}^n\), and \(\mathbf{I}_n\) denotes the identity matrix in \(\mathbb{R}^{n \times n}\). For any vector \(\mathbf{x}\), the matrix \(\text{diag}(\mathbf{x})\) refers to a diagonal matrix with the elements of \(\mathbf{x}\) on its main diagonal.

\begin{figure*}[h!]
    \centering
    \includegraphics[width=.8\linewidth]{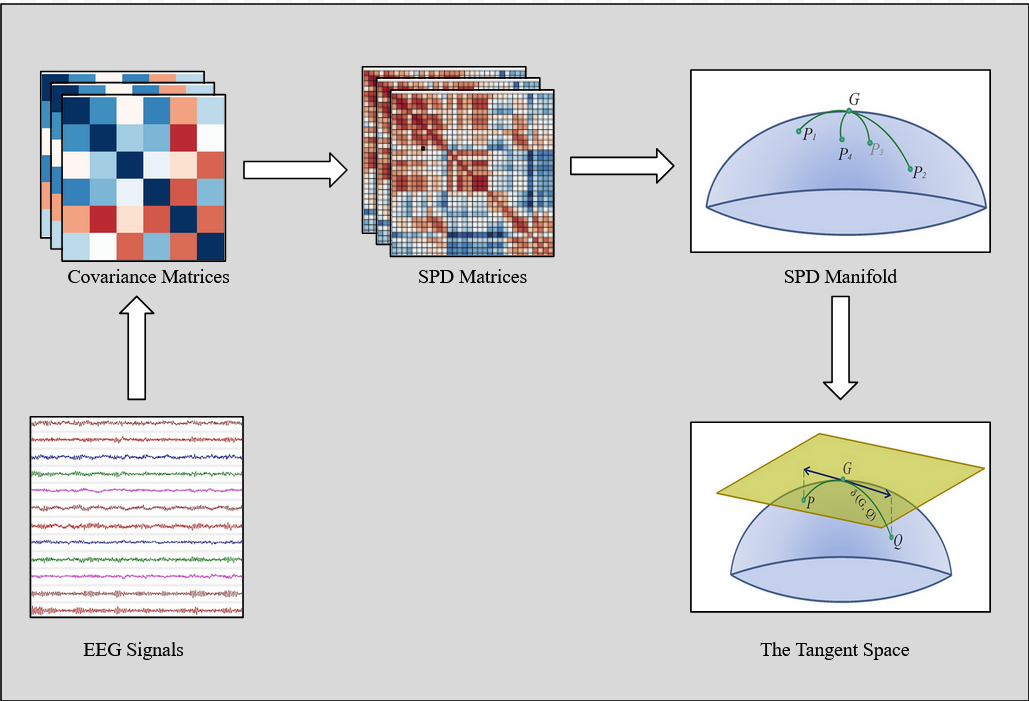}
    \caption{The SPD Manifold pipeline\cite{a76}}
    \label{fig:enter-label}
\end{figure*}
We denote by \(\text{vec}(\mathbf{A})\) the operation that converts a matrix \(\mathbf{A}\) into a vector by stacking its columns. In the special case where \(\mathbf{A}\) is symmetric, the vectorization \(\text{vec}(\mathbf{A})\) is defined such that it takes only the unique entries due to symmetry. The norms \(\|\cdot\|_p\) and \(\|\cdot\|_F\) correspond to the \(L_p\) and Frobenius norms of a vector or matrix, respectively.

\textbf{Definition 1.1.} A matrix \(\mathbf{A} \in \mathbb{R}^{n \times n}\) is deemed Symmetric Positive Definite (SPD) if it holds that \(\mathbf{A} = \mathbf{A}^T\) and \(\mathbf{x}^T\mathbf{A}\mathbf{x} > 0\) for all non-zero vectors \(\mathbf{x} \in \mathbb{R}^n\). The eigenvalues of such a matrix \(\mathbf{A}\), denoted by \(\lambda(\mathbf{A})\), are guaranteed to be positive.

\textbf{Definition 1.2.} A matrix \(\mathbf{A}\) is considered orthogonal if its columns comprise an orthogonal unit vector set, implying \(\mathbf{A}^T\mathbf{A} = \mathbf{I}_n\).

\textbf{Definition 1.3.} The exponential and logarithmic functions of a matrix \(\mathbf{A} \in \mathbb{R}^{n \times n}\), denoted by \(\exp(\mathbf{A})\) and \(\log(\mathbf{A})\), are defined through its eigenvalue decomposition. If \(\mathbf{A}\) is expressible as \(\mathbf{U}\text{diag}(\lambda_1, \ldots, \lambda_n)\mathbf{U}^T\), where \(\mathbf{U}\) is orthogonal and \(\lambda_i\) are the eigenvalues of \(\mathbf{A}\), then:
\begin{equation}
\exp(\mathbf{A}) = \mathbf{U}\text{diag}(\exp(\lambda_1), \ldots, \exp(\lambda_n))\mathbf{U}^T,
\end{equation}
\begin{equation}
\log(\mathbf{A}) = \mathbf{U}\text{diag}(\log(\lambda_1), \ldots, \log(\lambda_n))\mathbf{U}^T.
\end{equation}

A manifold, in the simplest terms, refers to a space that, at a local level, resembles the Euclidean space of dimension \( n \), known as \( \mathbb{R}^n \). This property of local similarity to \( \mathbb{R}^n \) qualifies it as a topological manifold \cite{a52}. When such a space is equipped with a differential structure, it ascends to become a differentiable manifold, thereby enabling the attachment of a tangent space at any point \cite{a53}. The tangent space encapsulates all potential vectors that are tangential to any conceivable curve passing through that point on the manifold.

Within the manifold paradigm, a Riemannian manifold stands out as a differentiable manifold that is further enhanced with a smoothly varying inner product on its tangent spaces, constituting the Riemannian metric tensor \cite{a54}. This metric tensor facilitates the measurement of angles between vectors within the same tangent space, the magnitude of vectors, and the distance between vectors. It allows for the calculation of curve lengths on the manifold, including the geodesic distance which is the shortest distance between two points on the manifold determined by the paths of curves \cite{a55}.

Considering \( \mathcal{M} \) as a Riemannian manifold and \( T_X\mathcal{M} \) as its tangent space at point \( X \), for any two vectors \( V, W \in T_X\mathcal{M} \), the Riemannian metric \( g_{X}(V, W) \) provides the inner product \cite{a56}. When discussing the manifold of SPD matrices, our focus is particularly on the tangent space \( T_X\mathcal{M} \) and the Riemannian metric at that point.

For a smooth curve \( \gamma: [0,1] \rightarrow \mathcal{M} \), parametrized with respect to a metric on \( \mathcal{M} \), there are numerous ways to describe its trajectory, each with a different speed of traversal. Of particular interest are those curves that are parametrized by arc length, termed naturally parametrized curves \cite{a57}. These curves progress with a uniform rate and are defined in relation to the Riemannian metric. The curve's length from \( \gamma(0) \) to \( \gamma(1) \) is computed by integrating the metric-induced velocity along the curve.

For two distinct points \( X_1, X_2 \in \mathcal{M} \), and a naturally parametrized curve \( \gamma \) that connects these points with \( \gamma(0) = X_1 \) and \( \gamma(1) = X_2 \), the length \( L(\gamma) \) of this curve is expressed by the integral\cite{a52}:
\begin{equation}
    L(\gamma) = \int_{0}^{1} \sqrt{g_{\gamma(t)}(\dot{\gamma}(t), \dot{\gamma}(t))} \, dt 
\end{equation}

The optimal path on a manifold that minimizes the distance between two points is called a geodesic. While geodesics represent the paths of least distance, they are uniquely characterized by a uniform rate of traversal and are not necessarily the shortest paths for manifolds that are not simply connected. These paths, particularly on a spherical surface, might have multiple representations, such as the numerous geodesics connecting the poles \cite{a55}. The distance along a geodesic, known as the Riemannian distance between two points \( X_1 \) and \( X_2 \), is essential in determining the manifold's completeness. Here, we concentrate on manifolds that are geometrically complete \cite{a54}.

The tangent space at point \( X \) on a manifold \( \mathcal{M} \), denoted \( T_X\mathcal{M} \), acts as a local linear approximation to \( \mathcal{M} \) within a specific region around \( X \), usually within a certain radius allowing for a bijection via the exponential map \cite{a58}. For every point \( X' \) within this neighborhood, the exponential map is the bridge connecting \( X' \) to \( X \) through a unique geodesic.

For any smooth scalar function \( f \) defined on \( \mathcal{M} \), the Riemannian gradient at \( X \) is given as \( \nabla f(X) \) in the direction that maximizes the rate of increase of \( f \). If \( \gamma \) is a geodesic with \( \gamma(0) = X \) and \( \dot{\gamma}(0) = V \), then \( \nabla f \circ \gamma \) describes the change of \( f \) along \( \gamma \), and \( \nabla f(X) \) is that vector in \( T_X\mathcal{M} \) such that the inner product \( \langle V, \nabla f(X) \rangle \) equals the derivative of \( f \) along the curve at \( t=0 \), defined as\cite{a58}:
\begin{equation}
    \langle V, \nabla f(X) \rangle = \frac{d}{dt}f(\gamma(t))\bigg|_{t=0} .
\end{equation}

The Riemannian gradient facilitates the computation of directional derivatives on manifolds, linking classical differential calculus to the geometry of the manifold. The exponential map, denoted as \( \text{Exp}_X: T_X\mathcal{M} \rightarrow \mathcal{M} \), projects a tangent vector at \( X \) onto \( \mathcal{M} \), and conversely, the logarithm map, or \( \text{Log}_X \), maps points back to \( T_X\mathcal{M} \), providing a method to traverse between a tangent space and its manifold \cite{a59}.

Assigning a set \( \mathcal{M} \) of square matrices the structure of a Riemannian manifold incorporates a local Euclidean geometry, thereby enriching it with a rigorous mathematical underpinning. Suppose we have \( M' \in \mathcal{M} \), a manifold of dimension \( K \), and \( \xi_{M'} = M' - M \) representing a tangent vector at \( M \). This vector is part of a higher-dimensional tangent space \( T_{M} \mathcal{M} \) associated with \( M \).

The inner product defined by the Riemannian metric on \( T_{M} \mathcal{M} \times T_{M} \mathcal{M} \rightarrow \mathbb{R} \) induces a norm on the tangent space, given by \( \|\xi_{M'}\|_M = \sqrt{\langle \xi_{M'}, \xi_{M'} \rangle_M} \), and facilitates the computation of geodesic distances \( d(M, M') \) on \( \mathcal{M} \) \cite{a59}. These distances enable the formulation of mean values of points on the manifold as:
\begin{equation}
\text{Mean}(\{M_1, \dots, M_N\}) = \underset{M \in \mathcal{M}}{\arg\min} \sum_{i=1}^{N} d(M, M_i)^2.
\end{equation}

The exponential mapping at \( M \) in \( \mathcal{M} \), denoted as \( \text{Exp}_{M} \), and its inverse, the logarithm mapping \( \text{Log}_{M} \), are both essential in preserving the manifold's structure, especially when mapping to and from the tangent space. \( \text{Exp}_{M}(\xi_{M}) \) approximates \( M' \) when \( \xi_{M} \) is small, and the distance between \( M \) and \( M' \) is given by the norm of \( \xi_{M} \) in \( \mathbb{R}^K \) via \( \text{Log}_{M} \). This relationship allows the introduction of vectorization for \( \mathcal{M} \), \( \mathcal{P}_{M}: \mathcal{M} \rightarrow \mathbb{R}^K \), defined as \( \mathcal{P}_{M}(M') = \phi(\text{Log}_{M}(M')) \) \cite{a59}.

For a compact subset of \( \mathcal{M} \) where the matrices \( \{M_i\} \) lie, the mean \( M \) can be approximated, resulting in a simplified geodesic distance expression:
\begin{equation}
d(M, M') \approx \| \mathcal{P}_{M}(M) - \mathcal{P}_{M}(M') \|.
\end{equation}

In the realm of regression on manifolds, the vectorization \( \mathcal{P}_{M} \) is pivotal for leveraging machine learning algorithms. It adapts matrix points in \( \mathcal{M} \) onto \( \mathbb{R}^K \), facilitating their use in regression techniques that typically assume a Euclidean data structure. For instance, with a collection \( \{M_i\} \) and corresponding response variables \( \{y_i\} \), one first computes the mean of the samples \( M \) and applies vectorization to obtain \( \{v_i\} \). Linear regression methods, such as ridge regression, can then be applied, presupposing a linear relationship \( y_i \approx v_i^T \beta \), with \( \beta \) representing the regression coefficients \cite{a60}.

\subsection{The Covariance Matrix of EEG}

Consider \(\mathbf{X} \in \mathbb{R}^{M \times T}\), representing an EEG signal that has undergone band-pass filtering, where \(M\) denotes the number of channels and \(T\) represents the number of temporal samples. To analyze the statistical properties of the EEG signal, we construct the covariance matrix \(\mathbf{P}\) as follows:

\begin{equation}
\mathbf{P} = \frac{1}{T-1}\mathbf{X}\mathbf{X}^T
\end{equation}

This matrix \(\mathbf{P}\) is not only symmetric but also empirically confirmed to be SPD, encapsulating important statistical information about the EEG signals. The symmetry of \(\mathbf{P}\) arises because \((\mathbf{X}\mathbf{X}^T)^T = \mathbf{X}\mathbf{X}^T\), and it attains its positive-definiteness under the condition that \(\mathbf{X}\) has full row rank, which is typically satisfied if \(T > M\) and the EEG data are sufficiently diverse \cite{a61}.

\textbf{Properties of Symmetric Positive-Definite Matrices:}
A matrix \(\mathbf{A} \in \mathbb{R}^{n \times n}\) is deemed positive-definite if for any non-zero vector \(\mathbf{v} \in \mathbb{R}^n\), it holds that:

\begin{equation}
\mathbf{v}^T \mathbf{A} \mathbf{v} > 0
\end{equation}

Applying this to our covariance matrix \(\mathbf{P}\)\cite{a62}, we observe:

\begin{equation}
\mathbf{v}^T \mathbf{P} \mathbf{v} = \mathbf{v}^T \left(\frac{1}{T-1} \mathbf{X} \mathbf{X}^T\right) \mathbf{v} 
= \frac{1}{T-1} \|\mathbf{X}^T \mathbf{v}\|^2 > 0 
\end{equation}

where \(\|\mathbf{X}^T \mathbf{v}\|\) represents the Euclidean norm of \(\mathbf{X}^T \mathbf{v}\), which is greater than zero unless \(\mathbf{v}\) is orthogonal to all rows of \(\mathbf{X}\). This ensures that \(\mathbf{P}\) retains all the characteristics of an SPD matrix, including its utility in defining a metric space on the manifold of such matrices \cite{a63}.

\textbf{Geometric Interpretation in EEG Analysis:}
As an element of the manifold of SPD matrices, \(\mathbf{P}\) can be further analyzed using tools of Riemannian geometry. The diagonal elements of \(\mathbf{P}\) represent the variance of the filtered EEG signal at each channel, indicating the power spectrum. The off-diagonal elements provide a measure of covariance between different channels, reflecting how signal components vary together over time—a crucial aspect for understanding functional connectivity in the brain \cite{a64}. Distances between such matrices, computed via geodesic paths on this manifold, offer a natural measure of similarity between different EEG states or conditions, enhancing applications like classification and clustering in brain-computer interfaces \cite{a63}.

Incorporating these mathematical insights enriches our understanding of how the covariance matrix functions within EEG signal analysis and underscores the significance of employing a Riemannian framework to advance analytical capabilities.

\subsection{Riemannian Metrics on SPD Manifolds}

The choice of Riemannian metric on the manifold of SPD matrices significantly influences the analysis and processing of data represented by these matrices, such as EEG signals. The Log-Euclidean Metric (LEM) and the Affine-Invariant Metric (AIM) are two predominant metrics used in this context \cite{a63, a65}.

\textbf{Log-Euclidean Metric (LEM):}
The Log-Euclidean Metric (LEM) offers an efficient and robust computational method for handling the manifold of SPD matrices. This metric is particularly valued for preserving the bi-invariance property within the Lie group structure of SPD matrices, making it suitable for various applications, including image processing and medical imaging analysis, where maintaining the structure of data during transformations is crucial \cite{a65}.

Under the LEM, the geodesic distance between two points \(P_1\) and \(P_2\) on the manifold of SPD matrices is defined as:
\begin{equation}
    \delta_L(P_1, P_2) = \| \Log(P_1) - \Log(P_2) \|_F,
\end{equation}
where \(\Log\) denotes the matrix logarithm, transforming matrices to a space where the Euclidean tools can be applied. The norm \(\|\cdot\|_F\) represents the Frobenius norm, which measures matrix entries' absolute differences, thus providing a natural distance metric in the logarithmic domain \cite{a65}.

The Log-Euclidean mean of a set of matrices, crucial for statistical analysis on manifolds, is calculated using:
\begin{equation}
    G = \Exp\left( \frac{1}{k} \sum_{i=1}^{k} \Log(P_i) \right),
\end{equation}

This formulation allows the mean of matrices to be computed efficiently and without the convergence issues that may arise with other Riemannian metrics. Additionally, when considering a weighted mean where each matrix \(P_i\) is assigned a weight \(w_i\), fulfilling the conditions \(\sum_{i=1}^{k} w_i = 1\) and \(w_i > 0\), the weighted mean is given by:
\begin{equation}
    G = \Exp\left( \sum_{i=1}^{k} w_i\Log(P_i) \right).
\end{equation}

\textbf{Affine-Invariant Metric (AIM):}
The AIM is another highly regarded metric for SPD manifolds, especially valued for its invariant properties under affine transformations. The geodesic distance under this metric between \(P_1\) and \(P_2\) is computed as:
\begin{equation}
    \delta_A(P_1, P_2) = \left\| \Log(P_1^{-1/2} P_2 P_1^{-1/2}) \right\|_F,
\end{equation}
reflecting the minimal invariant distance under congruent transformations \cite{a63}. The mean under AIM, often used for central tendency estimation in statistics on manifolds, does not have a closed-form solution and is typically computed numerically \cite{a66}.

\textbf{Other Important Riemannian Metrics:}

\textbf{1. Fisher-Rao Metric:}
Originally developed in the context of information geometry, the Fisher-Rao metric has been adapted to the geometry of SPD manifolds. It provides a statistical interpretation of the distance, correlating with the intrinsic curvature of the data space, making it suitable for probabilistic interpretations of data \cite{a67}.

\textbf{2. Cholesky Decomposition-based Metric:}
This metric leverages the Cholesky decomposition of SPD matrices, enabling an alternative parametrization of the manifold. It simplifies certain computational procedures, particularly in optimization contexts, by linearizing the manifold using lower triangular matrices \cite{a68}.

\textbf{3. Procrustes Metric:}
Focused on shape analysis, the Procrustes metric measures the similarity between shapes after transformations like translation, scaling, and rotation. When adapted to SPD matrices, it helps analyze shape changes in data structures characterized by these matrices \cite{a60}.

Each of these metrics offers unique advantages depending on the specific requirements of the application, such as computational efficiency, robustness to transformations, or interpretability in statistical analysis. Choosing the appropriate metric is crucial for effectively analyzing and interpreting the geometric structure of data residing on SPD manifolds.

\subsection{Tangent Space at a Point on a Manifold}

The concept of tangent space is central to understanding the local geometry of manifolds and plays a critical role in the analysis of data that lie on these manifolds, such as covariance matrices of EEG signals. A tangent space at a point on a manifold provides a linear approximation of the manifold near that point, facilitating operations like vector addition and scalar multiplication which are not inherently defined on the manifold itself \cite{a52}.

Let \(\mathcal{M}\) be a smooth manifold and \(p\) a point on \(\mathcal{M}\). The tangent space to \(\mathcal{M}\) at \(p\), denoted as \(T_p\mathcal{M}\), is a vector space consisting of the tangent vectors to all possible curves through \(p\) on the manifold. Formally, if \(\gamma : (-\epsilon, \epsilon) \to \mathcal{M}\) is a smooth curve with \(\gamma(0) = p\), then the derivative \(\gamma'(0)\), which represents the velocity vector of \(\gamma\) at \(p\), is a tangent vector at \(p\) \cite{a53}.

For practical computation, particularly in applications involving SPD matrices, we express tangent vectors in coordinates. If \(\mathcal{M}\) is parameterized locally around \(p\) by a coordinate system \(\phi : U \subset \mathbb{R}^n \to \mathcal{M}\), where \(U\) is an open set in \(\mathbb{R}^n\), the tangent vectors can be expressed as:
\begin{equation}
\mathbf{v} = \sum_{i=1}^n v^i \frac{\partial}{\partial x^i}\bigg|_p
\end{equation}
where \(v^i\) are components of \(\mathbf{v}\) in the coordinate basis \(\left\{ \frac{\partial}{\partial x^i} \bigg|_p \right\}\) \cite{a52}.

In the context of SPD matrices, which form an open subset of the space of symmetric matrices, the tangent space at any point \(P \in \mathcal{SPD}(n)\) can be identified with the space of symmetric matrices. If \(P\) is an SPD matrix and \(\mathbf{S}\) is a symmetric matrix, then a curve on \(\mathcal{SPD}(n)\) passing through \(P\) with direction \(\mathbf{S}\) at \(t = 0\) can be represented as:
\begin{equation}
\gamma(t) = P^{1/2} \exp(t P^{-1/2} \mathbf{S} P^{-1/2}) P^{1/2}
\end{equation}
where \(\exp\) denotes the matrix exponential. The derivative of this curve at \(t = 0\) gives a tangent vector at \(P\), which is precisely \(\mathbf{S}\) \cite{a63}.

Understanding the tangent space of SPD matrices is crucial for developing algorithms in EEG signal processing. For example, the process of projecting EEG covariance matrices onto the tangent space allows for the linearization of the manifold structure, simplifying computations such as averaging or classification. This projection involves mapping SPD matrices to their tangent spaces at a chosen reference point, usually the mean covariance matrix, and performing linear operations in this vector space before projecting back to the manifold \cite{a69}.
\begin{figure}[h!]
    \centering
    \includegraphics[width=.7\linewidth]{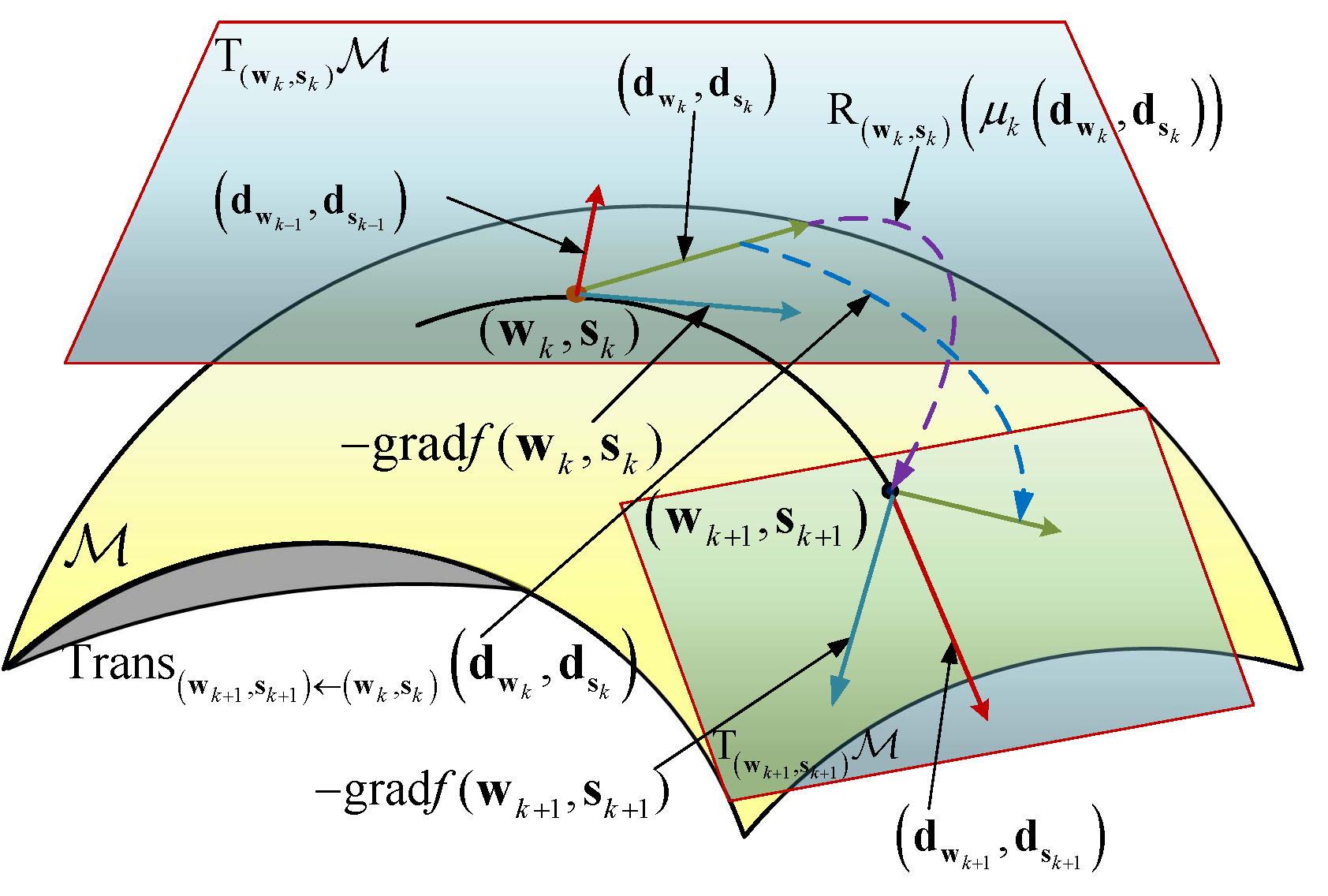}
    \caption{Illustration of the gradient descent algorithms on the Riemannian product manifold\cite{b49}. }
    \label{fig:enter-label}
\end{figure}
\subsection{Geodesic Distances on Riemannian Manifolds}

An SPD matrix of size \(M \times M\) resides on a manifold denoted as \(\mathcal{P}(M)\), characterized by its smooth and curved structure. On such Riemannian manifolds, the concept of geodesics plays a crucial role in understanding and quantifying the intrinsic geometric properties of the space \cite{a70}.

Geodesics are the generalization of "straight lines" to Riemannian manifolds and represent the shortest path between two points on the manifold. Mathematically, a geodesic \(\gamma(t)\) on a manifold \(\mathcal{M}\) satisfies the geodesic equation:
\begin{equation}
\nabla_{\dot{\gamma}(t)} \dot{\gamma}(t) = 0,
\end{equation}
where \(\nabla\) denotes the Levi-Civita connection associated with the manifold's Riemannian metric, and \(\dot{\gamma}(t)\) is the tangent vector to \(\gamma\) at \(t\). This equation ensures that the acceleration of \(\gamma\) is zero, signifying that the path has no deviation other than what is necessary to remain on the manifold \cite{a54}.

To compute a geodesic path under the Log-Euclidean framework, one uses the exponential and logarithmic maps. For matrices \(\mathbf{P}_1\) and \(\mathbf{P}_2\), the geodesic connecting them can be parameterized as:
\begin{equation}
\gamma(t) = \Exp\left((1-t) \Log(\mathbf{P}_1) + t \Log(\mathbf{P}_2)\right)
\end{equation}
for \(t \in [0,1]\), where \(\Exp\) is the matrix exponential. This expression provides an explicit parametric form for the geodesic, connecting two points on the manifold in a manner that respects the curvature and topological constraints \cite{a65}.

The curvature of the manifold, which affects the shape of geodesics, can be quantified through the Riemann curvature tensor \(R\), defined by:
\begin{equation}
R(X, Y)Z = \nabla_X \nabla_Y Z - \nabla_Y \nabla_X Z - \nabla_{[X, Y]}Z
\end{equation}
for vector fields \(X, Y,\) and \(Z\). The curvature informs us about how the metric space bends locally and impacts geodesic deviation \cite{a70}.

The ability to measure and analyze distances between EEG covariance matrices using geodesics\cite{a71} is invaluable for tasks such as classification and clustering within brain-computer interfaces\cite{a72}. Understanding the manifold's geometric structure\cite{a73} allows for more nuanced insights into brain connectivity patterns and functional dynamics, facilitating advanced neuroscientific analyses\cite{a74}\cite{a75}.

\section{Riemannian Geometry-Based EEG Approaches}
This article presents a synthesis of discoveries from a meticulously curated collection of 42 publications, manually chosen based on their pertinence to the fusion of Riemannian geometry and deep learning in brain-computer interfaces (BCIs). The selection of articles was made through a focused exploration conducted on Google Scholar and PubMed. The exploration, executed with the objective of encompassing the most recent and pertinent research up to May 3rd, 2024, utilized a search query formulated as: (“Riemannian geometry” OR “manifold learning” Or “SPD Manifolds”) AND (“brain computer interface” OR  “EEG”). This curation procedure encompassed both peer-reviewed publications and preprints, ensuring a thorough incorporation of the newest advancements and dialogues in the domain. 

To furnish a structured and comprehensive analysis of the vast body of literature concerning the utilization of Riemannian geometry in brain-computer interfaces, the chosen studies have been classified into five distinct domains: Feature Extraction, Classification, Manifold Learning, Tangent Space Methods, and Transfer Learning. This classification was selected to mirror the foundational phases of BCI signal processing and the specific obstacles that each phase poses. The core elements of BCI systems are encapsulated in feature extraction and classification, with a focus on extracting meaningful patterns from raw EEG data and subsequently utilizing these patterns for making predictions or decisions. Manifold learning and tangent space methods are segregated due to their significance in addressing the intrinsic non-Euclidean nature of EEG data, enabling a more intricate examination of how Riemannian geometry can enhance these procedures. Notably, transfer learning is delineated as a distinct category owing to its escalating importance in augmenting the adaptability and efficiency of BCIs across various subjects and sessions, showcasing the pragmatic implications of Riemannian techniques in real-world scenarios.

\begin{table*}[ht]
\centering
\caption{Overview of the reviewed related works}
\label{your-label-here}
\begin{tabularx}{.9\textwidth}{ll} 
\toprule
Category                   & Papers \\ \midrule
Feature Extraction    & \cite{b28},\cite{b33},\cite{b14},\cite{b41}\\
Classification Approaches              & \cite{b18},\cite{b4},\cite{b3},\cite{b22},\cite{b9},\cite{b47},\cite{b46},\cite{b1},\cite{b26},\cite{b25},\cite{b15},\cite{b40},\cite{b37} \\
Manifold Learning       & \cite{b7},\cite{b6},\cite{b10},\cite{b11},\cite{b20},\cite{b10},\cite{b21} ,\cite{b39},\cite{b48},\cite{b24},\cite{b23},\cite{b30}\\
Tangent Space Approaches         & \cite{b32},\cite{b17},\cite{b16},\cite{b19},\cite{b5},\cite{b43},\cite{b2} \\
Transfer Learning Approaches  & \cite{b8},\cite{b34},\cite{b31},\cite{b13},\cite{b38} \\ \bottomrule
\end{tabularx}
\end{table*}

\subsection{Feature Extraction Methods}
Feature extraction is a SOTA necessary step after collecting EEG data. It allows reducing the dimensionality of the data and making it more comprehended by machine learning and deep learning models. Various studies have explored the use of Riemannian geometry for feature extraction and discriminant analysis in EEG-based BCIs. In \cite{b28}, the authors used Covariance Matrices as EEG Signal Descriptors and exploring various dissimilarity metrics on SPD matrices and introducing a novel feature, HORC, which combines different relevance matrices under a tensor framework for improved classification accuracy. A Riemannian Spatial Pattern (RSP) method was proposed in \cite{b33} to extract spatial patterns from Riemannian classification for motor imagery tasks. The RSP method uses a backward channel selection procedure and compares it with the Common Spatial Pattern (CSP) approach. The RSP method provides precise mapping and clustering of imagined motor movements, which is especially useful in differentiating finger flexions. The spatial pattern extraction can be described using:

\begin{equation}
C = \frac{1}{N} \sum_{i=1}^{N} x_i x_i^T
\end{equation}

where $C$ is the covariance matrix, and $x_i$ represents the EEG signals.

Huang et al. \cite{b14} introduce the Common Amplitude-Phase Measurement (CAPM) method, designed to simultaneously analyze the amplitude and phase information of EEG signals on a Riemannian manifold. This dual consideration promises to enhance classification accuracy for BCI applications significantly.
The initial step involves extracting the amplitude and phase from EEG signals using the wavelet transform, specifically applying a complex Morlet wavelet. The transformation is mathematically represented as:
\begin{equation}
\hat{x}_c(t, f) = x_c(t) * \psi(t, f) = a_c(f, t)e^{i\theta_c(f,t)}
\end{equation}
where $\hat{x}_c(t, f)$ denotes the complex signal representation at channel $c$, time $t$, and frequency $f$. Here, $a_c(f, t)$ is the amplitude, $\theta_c(f, t)$ represents the phase, and $\psi(t, f)$ is the Morlet wavelet function.
To dimensionally reduce while preserving discriminative features, the authors devise a Riemannian graph embedding technique. The adjacency matrix $S$, based on the Riemannian distance between the SPD covariance matrices of EEG trials, is defined by:
\begin{equation}
s_{nr} = 
\begin{cases} 
\exp\left(-\frac{\delta_{nr}}{\sigma}\right), & \text{if } Y_n \in C_z(Y_r) \text{ and } Y_r \in C_z(Y_n) \\ 
0, & \text{otherwise} 
\end{cases}
\end{equation}
where $\delta_{nr}$ is the Riemannian distance between covariance matrices $P_n$ and $P_r$:

\begin{equation}
\delta_{nr} = \delta_R(P_n, P_r) = \left\| \log(P_n^{-1} P_r) \right\| = \left( \sum_{l=1}^L \log^2 \beta_l \right)^{1/2}
\end{equation}

The low-dimensional data $\hat{Y}_n$ is used to compute a new covariance matrix $\hat{P}_n = W^H P_n W$. Classification is performed using the Minimum Distance to Riemannian Mean (MDRM):

\begin{equation}
z_n = \text{sign} \left[ \left( \sum_{m=1}^M \log^2 \eta_m \right)^{1/2} - \left( \sum_{m=1}^M \log^2 \rho_m \right)^{1/2} \right]
\end{equation}

where $\eta_m$ and $\rho_m$ are the eigenvalues of $\hat{P}_{-1}^{-1} \hat{P}_n$ and $\hat{P}_{+1}^{-1} \hat{P}_n$ respectively. A regularized linear regression model is then applied to optimize the classification parameter $b$:

\begin{equation}
\min_b \frac{1}{2} \left\| z - Db \right\|^2 + \lambda (1 - \alpha) \left\| b \right\|_2^2 + \lambda \alpha \left\| b \right\|_1
\end{equation}

where $D$ is the matrix of input vectors $d_n$ and $z$ is the corresponding label vector.

The CAPM method effectively captures the intrinsic amplitude and phase information of EEG signals, optimizing the spatial-spectral filters and enhancing classification performance through Riemannian geometry-based regularization. Experimental results demonstrate significant improvements in classification accuracy on BCI competition datasets.

Gurve et al. \cite{b41} proposed a framework for classifying motor imagery EEG data using covariance matrices as descriptors and investigating various dissimilarity metrics on the manifold of SPD matrices. The study compares the performance of Log-Euclidean distance, Stein divergence, Kullback–Leibler divergence, and Von Neumann divergence for classification. Additionally, the paper introduces a new feature, Heterogeneous Orders Relevance Composition (HORC), combining different relevance matrices (Covariance, Mutual Information, or Kernel Matrix) under a tensor framework and multiple kernel fusion. The framework is further refined using Neighborhood Component Feature Selection (NCFS) to optimize the feature subset.

\cite{b41} also introduced a method to improve MI classification performance by employing Non-negative Matrix Factorization (NMF) for EEG channel selection and using the Riemannian geometry of covariance matrices for feature extraction. The method reduces the dimensionality of the EEG data, mitigates overfitting, and enhances classification accuracy by selecting subject-specific channels. The NMF is used to decompose the covariance matrix as follows:

\begin{equation}
C \approx WH
\end{equation}

where $C$ is the covariance matrix, and $W$ and $H$ are non-negative matrices. The neighborhood component feature selection (NCFS) algorithm is then applied to select the most important features, further refined by:

\begin{equation}
Dw(f_i, f_j) = \sum w_k^2 |f_{ik} - f_{jk}|
\end{equation}

where $w_k$ is the weighting vector for the $k$-th feature.

These studies collectively advance the application of Riemannian geometry in EEG feature extraction and discriminant analysis, offering improved accuracy, robustness, and computational efficiency in BCI systems.

\subsection{Classficiation Approaches}
Recent approaches of BCI classification  have notably shifted from traditional Euclidean metrics to employing Riemannian geometry, better reflecting the complex data structures. \cite{b18} and \cite{b4} demonstrated the use of Riemannian distance-based kernels within SVMs to significantly enhance classification accuracy for motor imagery tasks, surpassing traditional methods without extensive spatial filtering. Similarly, \cite{b3} documents substantial performance improvements in steady-state visually evoked potential (SSVEP) classification by aligning methods more closely with intrinsic data geometry. The integration of Riemannian geometry with decision tree frameworks in \cite{b22} improves classification by capturing non-linear data relationships. Additionally, \cite{b9} introduces an expectation-maximization algorithm for robust covariance estimation in the presence of incomplete EEG data, outperforming traditional imputation techniques. Furthermore, \cite{b47} develops new Riemannian geometry-based metrics to monitor and enhance user performance during BCI training, ensuring a more accurate and reliable progress assessment.

The work \cite{b46} explores the application of graph neural networks (GNNs) on SPD manifolds to classify motor imagery  EEG signals. This approach leverages the time-frequency characteristics of EEG data, which are represented on SPD manifolds. The mathematical formulation involves constructing a graph where each node represents an EEG channel, and edges are weighted by a function of the Riemannian distance between SPD matrices. The SPD matrices are derived from the covariance of time-frequency representations of the EEG signals, capturing both spatial and spectral information.

The key mathematical components include the computation of the Riemannian distance between two SPD matrices \( \Sigma_1 \) and \( \Sigma_2 \):

\begin{equation}
\delta_R(\Sigma_1, \Sigma_2) = \left\| \log(\Sigma_1^{-1/2} \Sigma_2 \Sigma_1^{-1/2}) \right\|_F,
\end{equation}

where \( \log \) denotes the matrix logarithm and \( \left\| \cdot \right\|_F \) is the Frobenius norm. This distance metric is crucial for defining the weights of the edges in the graph neural network, ensuring that the intrinsic geometry of the data is preserved.

The GNN model processes the SPD matrices using graph convolutional layers specifically designed for manifold-valued data. Let \( \mathcal{G} = (\mathcal{V}, \mathcal{E}) \) be a graph with vertices \( \mathcal{V} \) corresponding to EEG channels and edges \( \mathcal{E} \) weighted by the Riemannian distance. The signal at each vertex \( v \) is represented by an SPD matrix \( \Sigma_v \). The graph convolution operation on SPD manifolds is defined as:

\begin{equation}
H^{(l+1)} \leftarrow \text{RBN}\left(\text{ReEig}\left(W^{(l)}\left(\tilde{D}^{-1}\tilde{A}\right)H^{(l)}W^{(l)T}\right)\right)
\end{equation}

where \( \mathbf{H}^{(l)} \) is the feature matrix at layer \( l \), \( \mathbf{W}^{(l)} \) is the trainable weight matrix, \( \mathcal{N}(v) \) denotes the neighbors of vertex \( v \), and \( \sigma \) is a non-linear activation function. This operation ensures that the convolution respects the manifold structure of the data.

\cite{b37} describes a method that combines Independent Component Analysis (ICA) with Riemannian geometry to enhance emotion recognition from EEG signals; this involves projecting covariance matrices onto the Riemannian manifold and integrating these features into a deep learning model, resulting in superior performance compared to traditional methods. Similarly, \cite{b1} applies Riemannian geometry decoding algorithms to large-scale EEG datasets, where the baseline minimum distance to Riemannian mean approach yields the highest classification accuracy for motor imagery and execution tasks, underscoring the scalability of these methods. Additionally, \cite{b26} explores the feasibility of imagined speech classification using EEG signals, employing covariance matrix descriptors on the Riemannian manifold with a relevance vector machine classifier to achieve high accuracy, revealing promising potential for BCI applications in speech imagery.

In the paper \cite{b25} the authors propose a novel brain-ventilator interface (BVI) framework that detects patient-ventilator disharmony from EEG signals. This work leverages the spatial covariance matrices of EEG signals and utilizes Riemannian geometry to classify different respiratory states. The approach is robust against the non-stationarity and noise inherent in EEG signals. Mathematically, the framework involves calculating the spatial covariance matrix \( C \) for the EEG signals, which is SPD. The Riemannian mean \( C_k \) of these matrices is computed using:
\begin{equation}
C_k = \arg_{C} \min \sum_{C_i\in S_k} d(C, C_i)
\end{equation}

The features are then projected onto the tangent space using the logarithmic map:
\begin{equation}
\log_{Q}(P) = S_Q = Q^{1/2} logm(Q^{{-1/2}} P Q^{1/2}) Q^{1/2}
\end{equation}

These features are used in a classifier to detect respiratory states, thus enabling the BVI system.

The work on \cite{b15} introduces an innovative method combining Riemannian geometry with sparse optimization and Dempster-Shafer theory for enhanced motor imagery classification. The method, known as RSODSF, extracts features by first calculating the covariance matrices from segmented EEG signals, projecting them into Riemannian tangent space, and applying sparse optimization:
\begin{equation}
w = \arg \min_{w} \frac{1}{2} \left\| y - Fw \right\|_2^2 + \lambda \left\| w \right\|_1
\end{equation}
where \( F \) is the matrix of features, \( y \) is the label vector, \( w \) is the weight vector, and \( \lambda \) is the regularization parameter. The probabilistic outputs of a support vector machine (SVM) classifier are fused using Dempster-Shafer theory to improve classification accuracy:
\begin{equation}
m_{1,2}(EA) = \begin{cases} 
\frac{\sum_{EB \cap EC = EA} m_1(EB) \times m_2(EC)}{1 - \sum_{EB \cap EC = \emptyset} m_1(EB) \times m_2(EC)} & \text{if } EA \neq \emptyset \\
0 & \text{if } EA = \emptyset
\end{cases}
\end{equation}
The proposed method demonstrates significant accuracy improvements on BCI competition datasets.

The paper \cite{b40} suggested an expansion of the generalized learning vector quantization (GLVQ) to the curved Riemannian manifold of SPD matrices. The scholars introduce a technique known as Generalized Learning Riemannian Space Quantization (GLRSQ), which adjusts the GLVQ framework to operate within the suitable Riemannian metric, thereby notably improving classification accuracy for tasks involving EEG-based motor imagery.
The cost function for GLRSQ on a Riemannian manifold is given by:
\begin{equation}
E(W) = \sum_{i=1}^{m} \phi\left(\mu(X_i, W)\right)
\end{equation}
where \(\phi\) is a monotonically increasing function, and \(\mu(X_i, W)\) is defined as:
\begin{equation}
\mu(X_i, W) = \frac{\delta(X_i, W_J) - \delta(X_i, W_K)}{\delta(X_i, W_J) + \delta(X_i, W_K)}
\end{equation}
with \(W_J\) and \(W_K\) being the closest correct and incorrect prototypes, respectively.
The gradients of the cost function with respect to the prototypes are derived as:
\begin{equation}
\begin{aligned}
\Delta W_J &= -\alpha(t) \nabla_{W_J} E = -\alpha(t) \phi' \frac{4 \delta_K}{(\delta_J + \delta_K)^2} \text{Log}_{W_J}(X_i), \\
\Delta W_K &= \alpha(t) \nabla_{W_K} E = \alpha(t) \phi' \frac{4 \delta_J}{(\delta_J + \delta_K)^2} \text{Log}_{W_K}(X_i),
\end{aligned}
\end{equation}
where \(\alpha(t)\) is the learning rate, \(\phi'\) is the derivative of \(\phi\), and \(\text{Log}_{W_J}(X_i)\) denotes the logarithmic map of \(X_i\) at \(W_J\).
The exponential and logarithmic maps are used to update the prototypes on the SPD manifold:
\begin{equation}
\begin{aligned}
\text{Exp}_{W}(V) &= W^{1/2} \exp(W^{-1/2} V W^{-1/2}) W^{1/2}, \\
\text{Log}_{W}(X) &= W^{1/2} \log(W^{-1/2} X W^{-1/2}) W^{1/2}.
\end{aligned}
\end{equation}
The update rules for the prototypes \(W_J\) and \(W_K\) are given by:
\begin{equation}
\begin{aligned}
W_J &\leftarrow \text{Exp}_{W_J} \left(-\alpha(t) \nabla_{W_J} E\right), \\
W_K &\leftarrow \text{Exp}_{W_K} \left(-\alpha(t) \nabla_{W_K} E\right).
\end{aligned}
\end{equation}

This innovative approach ensures that the prototypes remain within the SPD manifold, leveraging the geometric properties of the manifold to enhance classification performance. The experimental results demonstrate that GLRSQ significantly outperforms traditional Euclidean-based GLVQ methods and shows competitive performance with state-of-the-art techniques in EEG classification for motor imagery tasks.

\subsection{manifold learning}
Integration of deep learning with Riemannian geometry-based methods was a tren in recent years. In \cite{b7}, the proposed EEG-SPDNet incorporates Riemannian geometry into deep network architectures, enhancing the decoding by exploiting the physiological plausibility of frequency regions. Concurrently, \cite{b6} introduces a method for robust representation of EEG signals through spatial covariance matrices, capturing homogeneous segments effectively. Furthermore, \cite{b10} develops a model that optimizes subject-specific frequency band selection for motor imagery classification by constructing multiple Riemannian graphs and applying advanced graph embedding and fusion techniques, tailoring the approach to individual variations in EEG signal patterns.

\cite{b11} introduces a novel method to address the limitations of existing SPD matrix-based Riemannian networks. They propose Riemannian Embedding Banks (REB), which partition the problem of learning common spatial patterns into multiple subproblems, each modeled separately and then combined within SPD neural networks. The REB method utilizes the SPDNet framework, which includes layers such as BiMap and ReEig for transformation and feature extraction. 
The BiMap layer performs a bilinear mapping to generate more discriminative and compact SPD matrix features. The transformation is given by:
\begin{equation}
X_k = W_s X_{s-1} W_s^T,
\end{equation}
where \( W_s \) is the transformation matrix ensuring the output \( X_k \) remains in the form of an SPD matrix.

Similar to the ReLU layer in traditional neural networks, the ReEig layer rectifies SPD matrices with small positive eigenvalues:
\begin{equation}
X_s = U_{s-1} \max(\epsilon I, \Sigma_{s-1}) U_{s-1}^T,
\end{equation}
where \( U_{s-1} \) and \( \Sigma_{s-1} \) are obtained from the eigenvalue decomposition \( X_{s-1} = U_{s-1} \Sigma_{s-1} U_{s-1}^T \), and \( \epsilon \) is a threshold parameter.
The REB approach optimizes the embedding by assigning features to clusters and ensuring that the samples within each cluster are informative. The clustering is achieved by minimizing:
\begin{equation}
L_A(X_i; f_p) = \sum_{X_j \in f_p(X_i)} \delta_R^2(X_i, X_j),
\end{equation}
\begin{equation}
L_P(X_i, y_i; f_p) = -\sum_{X_j \in f_p(X_i), y_j = y_i} \delta_R^2(X_i, X_j),
\end{equation}
and
\begin{equation}
L_N(X_i; f_p) = \sum_{X_m \in f_p(X_i), y_m \neq y_i} \log(\min(10^{-4}, Q(D(X_i, X_m))),
\end{equation}
where \( Q(D) \) is a distribution function ensuring negative samples are uniformly distributed based on their distances.
The final embedding is constructed by concatenating the sub-embeddings produced by individual learners:
\begin{equation}
f = [f_d(X_i; f_1), f_d(X_i; f_2), \ldots, f_d(X_i; f_K)],
\end{equation}
and the loss function for training is given by:
\begin{equation}
L_k = \lambda_1 \sum_{X_i \in C_k} L_{C_k}(X_i; f_p) + \lambda_2 \sum_{X_i \in C_k} L_{f_k}(X_i; f_k).
\end{equation}

The experimental results on public EEG datasets demonstrate the superiority of the proposed REB approach in learning common spatial patterns of EEG signals, increasing convergence speed, and maintaining generalization despite the non-stationary nature of the data.

The paper \cite{b20} leverages the graph structure of EEG trials, incorporating node influence properties and an ensemble of semantic graphs into the neural structured learning framework. This approach improves classification performance by capturing the complex relationships between EEG trials and integrating manifold regularities into the learning process. The optimization equation includes supervised and neighbor loss functions:
\begin{equation}
\text{loss} = \sum_{i=1}^D L(y_i, \hat{y}_i) + \alpha \sum_{i=1}^D L_N(y_i, x_i, N(x_i)),
\end{equation}
where \(L(y_i, \hat{y}_i)\) is the supervised loss, \(L_N(y_i, x_i, N(x_i))\) is the neighbor loss, and \(\alpha\) is the graph regularization parameter.

\cite{b42} presents a novel approach using geodesic correlation on the SPD manifold. The geodesic correlation measure $G(X, Y)$ for two SPD matrices $X$ and $Y$ is given by:

\begin{equation}
G(X, Y) = \sum_{i=1}^{d} \log^2 \lambda_i
\end{equation}

where $\lambda_i$ are the generalized eigenvalues of the pair $(X, Y)$. This framework enhances geodesic correlation for multi-view, self-supervised learning.

\cite{b10} proposes the MRGF model to optimize frequency band selection and extract spatial and spectral features from EEG signals. The framework constructs multiple Riemannian graphs and employs graph embedding and fusion techniques. The bilinear mapping $W$ for dimensionality reduction is optimized by preserving distance structures between the high-dimensional manifold and low-dimensional embedding:

\begin{equation}
\min_W \sum_{P_i, P_j \in C} \left| \delta_R(P_i, P_j) - \delta_R(W P_i W^T, W P_j W^T) \right|
\end{equation}

where $\delta_R$ is the Riemannian distance, and $C$ is the experimental dataset.

\cite{b21} proposes a novel manifold attention module designed to handle SPD matrices. This network employs bilinear mappings to convert input SPD matrices into query, key, and value matrices while retaining their SPD structure. Specifically, for a given input $\tilde{x}_i$, the query, key, and value mappings are formulated as follows:

\begin{equation}
q_i = h_q(\tilde{x}_i; W_q) = W_q \tilde{x}_i W_q^T
\end{equation}

\begin{equation}
k_i = h_k(\tilde{x}_i; W_k) = W_k \tilde{x}_i W_k^T
\end{equation}

\begin{equation}
v_i = h_v(\tilde{x}_i; W_v) = W_v \tilde{x}_i W_v^T
\end{equation}

where $W_q$, $W_k$, and $W_v$ are transformation matrices ensuring that the output remains in the SPD manifold. The similarity between $q_i$ and $k_j$ is computed using the Log-Euclidean distance:

\begin{equation}
\text{sim}(q_i, k_j) = \frac{1}{1 + \log(1 + \delta_1(q_i, k_j))} := \alpha_{ij}
\end{equation}

Here, $\delta_1$ denotes the Log-Euclidean distance. The attention matrix $\mathbf{A} = [\alpha_{ij}]_{m \times m}$ is normalized using the softmax function, and the final output $v_i'$ is computed as:

\begin{equation}
v_i' = \text{Exp}\left(\sum_{l=1}^{m} \alpha_{il} \text{Log}(v_l)\right)
\end{equation}

The backward procedure for gradient descent parameter updating on the Riemannian manifold involves projecting Euclidean gradients onto the tangent space using Stiefel gradients and retraction operations. The Stiefel gradient $\widetilde{\nabla}_{W_v}\mathcal{L}$ is computed as:

\begin{equation}
\widetilde{\nabla}_{W_v}\mathcal{L} = \nabla_{W_v}\mathcal{L} - \pi_N(\nabla_{W_v}\mathcal{L})
\end{equation}

\begin{equation}
\widetilde{\nabla}_{W_v}\mathcal{L} = \nabla_{W_v}\mathcal{L} - W_v\left(\frac{W_v^T\nabla_{W_v}\mathcal{L} + (\nabla_{W_v}\mathcal{L})^T W_v}{2}\right)
\end{equation}

Finally, the weight update is performed as:

\begin{equation}
W_v^{(new)} = \Gamma(W_v - \lambda \widetilde{\nabla}_{W_v}\mathcal{L})
\end{equation}

where $\Gamma$ is the retraction operation.

In \cite{b39}, the authors propose a novel method for modeling EEG covariance matrix distributions using Riemannian spectral clustering (RiSC). The framework allows for both unimodal and multimodal distributions on the SPD manifold, facilitating outlier detection and robust classification. The core of the method involves clustering EEG data points based on the Affine-Invariant Riemannian Metric (AIRM). 
For classification, the authors introduce the multimodal classifier mcRiSC, which computes the distance of a new observation \(P\) to cluster centroids and assigns it to the nearest cluster. The prediction is formulated as:
\begin{equation}
\hat{b} = \arg \min_{b \in \{1, \ldots, B\}} \delta_R \left( P, \bar{P}^{(b)} \right),
\end{equation}
where \(\bar{P}^{(b)}\) denotes the centroid of cluster \(b\). This method allows flexible application without prior knowledge of the data distribution, significantly enhancing classification performance.

The study on \cite{b48} demonstrates how projecting covariance matrices onto the Riemannian manifold, followed by their integration into a deep learning model, can yield superior performance compared to traditional methods. Similarly, the research on \cite{b24} investigates the application of geometric neural networks, leveraging phase space representations of EEG signals to improve classification accuracy. Furthermore, \cite{b23} employs a manifold-based GCN diffusion model to effectively capture complex spatial-temporal dynamics in EEG data, thereby enhancing the robustness and accuracy of BCI systems. Lastly, the work on \cite{b30} introduces an ensemble learning framework that combines Riemannian geometric features with machine learning classifiers, which significantly improves cross-session motor imagery classification accuracy.

These manifold learning approaches collectively advance the field of EEG signal processing by leveraging the mathematical properties of Riemannian manifolds to provide more effective and robust tools for feature extraction, data representation, and classification.

\subsection{Tangent Space Approaches}
Recent advancements in tangent space approaches for EEG signal processing have introduced innovative methods to enhance feature extraction and classification in BCIs. These methods leverage the properties of Riemannian geometry to map covariance matrices into the tangent space, allowing for effective manipulation and analysis of EEG data. \cite{b32} introduced a feature extraction method combining filter banks and Riemannian tangent space  for multi-category MI classification, addressing frequency variance and noise interference. \cite{b17} revisited Riemannian geometry-based EEG decoding through approximate joint diagonalization, simplifying Riemannian geometry concepts and reducing computational complexity. \cite{b16} proposed a Riemannian distance-based channel selection and feature extraction method combining discriminative time-frequency bands and Riemannian tangent space for MI-BCIs. \cite{b19} presented a spatio-temporal EEG representation learning framework on Riemannian manifold and Euclidean space, combining spatial and temporal information for improved classification performance.

\cite{b5} proposed a method for motor imagery classification that utilizes Riemannian geometry with a median absolute deviation (MAD) strategy. This approach involves calculating the average sample covariance matrices (SCMs) to select optimal reference metrics in a tangent space mapping (TSM)-based MI-EEG framework. The features are extracted using TSM, where the data from SCMs are projected according to a reference matrix representing the feature vector. Principal component analysis (PCA) and analysis of variance (ANOVA) are then applied to reduce dimensions and select optimal features for classification using linear discriminant analysis (LDA). The mathematical formulation of the Riemannian distance is given by:

\begin{equation}
d_R(C_i, C_j) = \left\| \log(C_i) - \log(C_j) \right\|_F
\end{equation}

where \( d_R \) represents the Riemannian distance, \( C_i \) and \( C_j \) are covariance matrices, and \( \left\| \cdot \right\|_F \) denotes the Frobenius norm. Miah et al. demonstrated that this method provides better accuracy compared to more sophisticated methods.

\cite{b43} introduced a manifold regression approach to predict from MEG/EEG signals without source modeling. This method uses the Riemannian geometry of rank-reduced covariance matrices and vectorizes them through projection into a tangent space. The Wasserstein distance and affine-invariant geometric distance are employed to handle rank-reduced data and improve prediction accuracy. The experimental results showed that this method outperforms sensor-space estimators and approaches the performance of source-localization models.

In detail, the covariance matrices of EEG signals are first computed as:

\begin{equation}
C_i = \frac{1}{N} \sum_{t=1}^{N} x_i(t) x_i(t)^T
\end{equation}

where \( C_i \) is the covariance matrix for the \(i\)-th trial, \( N \) is the number of time samples, and \( x_i(t) \) is the EEG signal at time \( t \).

These covariance matrices are then mapped to the tangent space of the Riemannian manifold. For a reference point \( C_0 \) on the manifold, the tangent space projection is given by:

\begin{equation}
\phi(C_i) = \log(C_0^{-1/2} C_i C_0^{-1/2})
\end{equation}

where \( \log \) is the matrix logarithm, and \( \phi(C_i) \) represents the projected covariance matrix in the tangent space.

For regression, the features are vectorized:

\begin{equation}
\phi(C_i) = \mathrm{vec}\left( \log(C_0^{-1/2} C_i C_0^{-1/2}) \right)
\end{equation}

where \( \mathrm{vec} \) denotes the vectorization operation.

The regression model is formulated as:

\begin{equation}
\hat{y} = W \cdot \phi(C)
\end{equation}

where \( \hat{y} \) is the predicted value, \( W \) is the weight matrix, and \( \phi(C) \) represents the feature vector obtained by projecting the covariance matrix \( C \) into the tangent space.

\cite{b2} explored the use of multiple tangent space projections for motor imagery EEG classification. The method involves projecting covariance matrices from their native Riemannian space to multiple class-dependent tangent spaces, enhancing the information provided to the classifier. This approach significantly improves model accuracy by utilizing different projections tailored to specific classes. The study demonstrated that multiple tangent space projections could effectively capture the discriminative features for better classification performance. The tangent space projection is mathematically expressed as:

\begin{equation}
\phi(C) = \mathrm{vec}\left( \log(C) - \log(C_0) \right)
\end{equation}

where \( \phi(C) \) is the feature vector, \( \log(C) \) is the matrix logarithm of the covariance matrix \( C \), and \( C_0 \) is the reference matrix.

These studies collectively advance the field of tangent space approaches for EEG signal processing, offering innovative solutions to enhance feature extraction, reduce dimensionality, and improve classification accuracy in BCIs.

\subsection{Transfer Learning Approaches}

Transfer learning approaches in EEG signal processing have made significant strides in reducing calibration time and improving classification accuracy by leveraging data from multiple subjects. These methods utilize Riemannian geometry and tangent space alignment to handle the variability between subjects and sessions effectively.

\cite{b8} proposed the Riemannian Geometric Instance Filtering (RGIF) framework to address the challenge of inter-subject variability in EEG signals. The framework includes two core components: a stable inter-subject similarity metric based on Riemannian geometry and a convolutional neural network for feature extraction. The similarity metric measures the similarity between a few trials from the target subject and abundant trials from source subjects, focusing on removing data from significantly different subjects. The geodesic distance between two SPD matrices \(P_1\) and \(P_2\) is defined as:

\begin{equation}
\delta_R(P_1, P_2) = \left\| \log(P_1^{-1}P_2) \right\|_F = \left( \sum_{i=1}^{n} \log^2 \lambda_i (P_1^{-1}P_2) \right)^{1/2}
\end{equation}

where \(\lambda_i\) are the eigenvalues of \(P_1^{-1}P_2\). The Riemannian mean matrix \(G_k\) for the target subject is computed as:

\begin{equation}
G_k = \argmin_{G} \sum_{i=1}^{N_k} \delta_R^2(G, P_i)
\end{equation}

This framework significantly reduces calibration time while maintaining high classification accuracy by combining data from similar subjects and the powerful feature extraction capabilities of CNNs.

\cite{b34} introduced supervised and semi-supervised manifold embedded knowledge transfer (sMEKT and ssMEKT) algorithms to handle inter-session/subject variability in MI EEG signals. These algorithms perform domain adaptation by preserving source domain discriminability and target domain geometric structure. After Riemannian alignment and tangent space mapping , both sMEKT and ssMEKT minimize the joint probability distribution shift between the source and target domains. The alignment process is given by:

\begin{equation}
\phi(C_i) = \log(C_0^{-1/2} C_i C_0^{-1/2})
\end{equation}

where \( \phi(C_i) \) represents the projected covariance matrix in the tangent space. The optimization for minimizing the distribution shift is formulated as:

\begin{equation}
\mathcal{L}(\theta) = \sum_{i=1}^{N_s} \ell(f_{\theta}(\phi(C_i^s)), y_i^s) + \lambda \sum_{i=1}^{N_t} \ell(f_{\theta}(\phi(C_i^t)), y_i^t)
\end{equation}

where \( \ell \) is the loss function, \( f_{\theta} \) is the classifier, \( C_i^s \) and \( C_i^t \) are source and target domain samples, respectively, and \( \lambda \) is a regularization parameter. These algorithms demonstrate significant improvements in classification accuracy with fewer labeled samples from the target domain.

Other notable contributions in transfer learning approaches include Xu et al. \cite{b13}, who proposed a selective cross-subject transfer learning approach based on Riemannian tangent space for motor imagery BCI, using Riemannian alignment to bring covariance matrices from different subjects closer and extracting Riemannian tangent space features for classification. Cai et al. \cite{b31} introduced a framework for motor imagery EEG decoding using manifold embedded transfer learning, aligning covariance matrices on the SPD manifold and extracting features from both SPD and Grassmann manifolds, demonstrating superior performance in transfer tasks. Additionally, \cite{b38} presented a tangent space alignment method for transfer learning in BCIs, aligning EEG data from different subjects in the tangent space of the positive definite matrices Riemannian manifold, showing improved classification accuracy. 

These studies collectively advance the field of transfer learning approaches for EEG signal processing, offering innovative solutions to enhance feature extraction, reduce calibration time, and improve classification accuracy in BCIs.

\section{Discussion and Challenges}

The integration of Riemannian geometry in EEG-based BCI systems has significantly progressed feature extraction, classification, and transfer learning within the respective fields. However, despite these advancements, various persistent challenges hinder the widespread adoption and efficacy of these methodologies. A primary obstacle lies in the computational intricacies associated with managing intricate manifold embeddings and high-dimensional transformations, particularly problematic in time-sensitive applications where swift data processing is essential. Furthermore, these approaches often struggle with generalization across diverse tasks and populations, given the variability in EEG patterns among individuals presents a notable hurdle. Additionally, the efficacy of Riemannian geometry-based techniques heavily relies on the quality of the input EEG data. Factors such as artifacts, noise, and non-stationarity can notably deteriorate performance, necessitating advanced preprocessing methods that may not always be practical in real-world scenarios. Another significant constraint is the lack of interpretability in these sophisticated mathematical transformations, which can impede clinical acceptance and further development of the methodologies. Moreover, there exists a noticeable gap in integrating these approaches with other modalities like fMRI or MEG, which could potentially enhance diagnostic and therapeutic capabilities but also introduce additional complexity in data fusion and analysis.
\begin{figure}[h!]
    \centering
    \includegraphics[width=\linewidth]{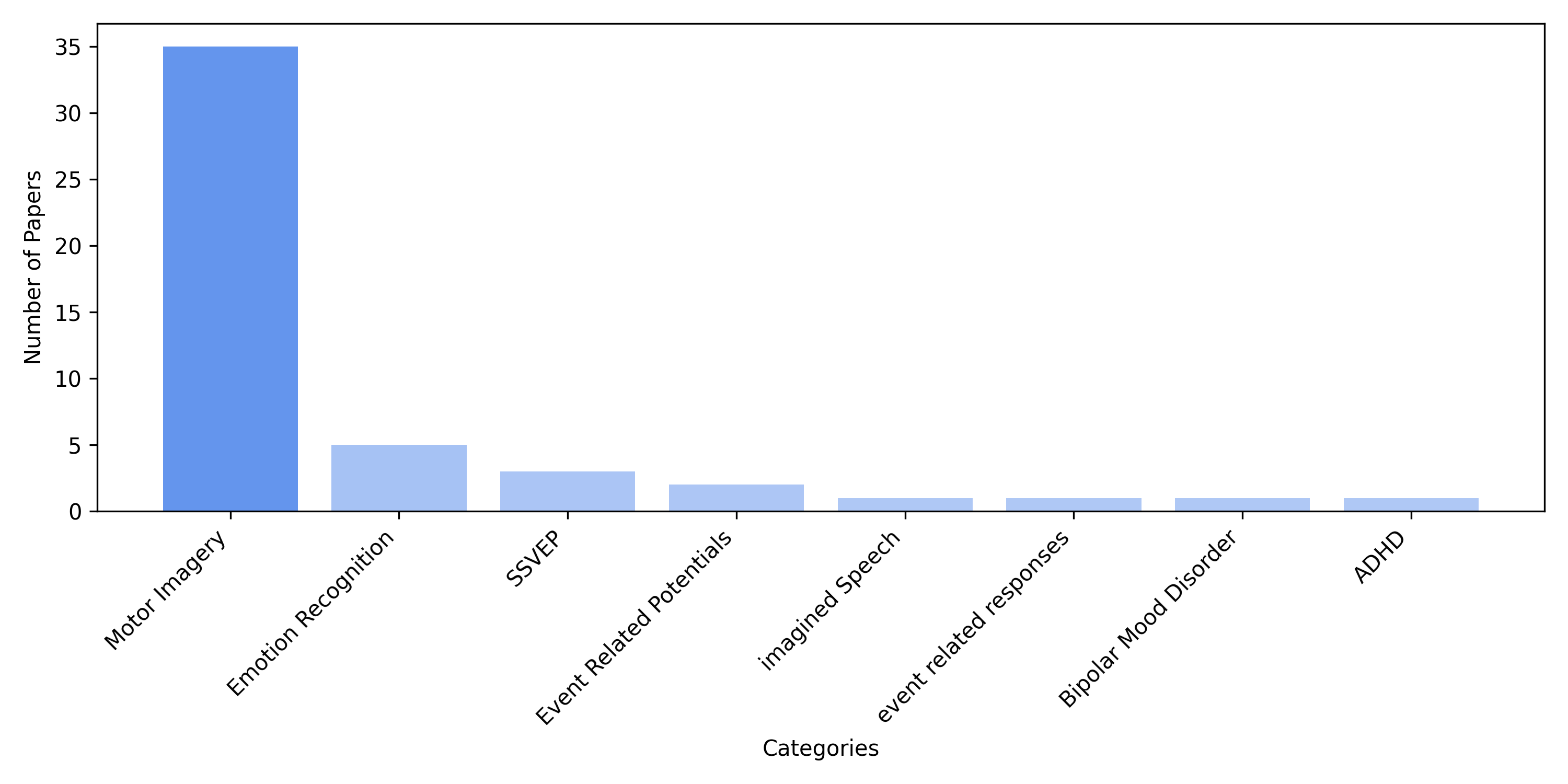}
    \caption{Classification tasks presented in the current literature }
    \label{fig:enter-label}
\end{figure}
Concentrated efforts in various crucial areas are imperative. Subsequent research endeavors should prioritize the enhancement of computational algorithms to achieve higher efficiency, thereby reducing the temporal and resource demands of existing methodologies, potentially through the utilization of emerging technologies like GPU acceleration. Additionally, there exists an urgent requirement for the creation of adaptive models capable of accommodating the inter-individual variabilities and cognitive state differences present across a heterogeneous population. Enhancements in managing noise and artifacts within the Riemannian framework have the potential to notably augment the quality of EEG data before further processing. The amalgamation of Riemannian geometry with alternative machine learning strategies in hybrid models could represent a significant advancement in enhancing both the efficacy and interpretability of these approaches. Furthermore, broadening the application of these methodologies to synergize with other neurological data modalities could offer a more comprehensive insight into brain function. Lastly, the establishment of open-source platforms and collaborative frameworks could expedite innovation and standardize practices in the field, thus rendering these sophisticated techniques more accessible and adaptable across various applications.

\subsection{Future Directions in Research}

To address the computational complexity of Riemannian geometry-based methods, future research should focus on optimizing algorithms for faster computation. This includes developing efficient numerical methods for calculating Riemannian distances and means, as well as exploring alternative representations of EEG signals that retain critical information with lower dimensionality. One potential direction is the use of approximation techniques or dimensionality reduction methods that preserve the geometric properties of the data.

Enhancing the scalability of graph neural networks on SPD manifolds is a critical area for prospective investigation. This task entails the creation of scalable GNN structures capable of effectively managing extensive EEG datasets. The exploration of hybrid models amalgamating the merits of conventional machine learning approaches and deep learning methodologies could contribute to tackling the scalability challenge. Moreover, delving into parallel computation and distributed processing methodologies might further amplify the scalability of these frameworks.

Adapting manifold learning strategies to render them more suitable for real-time applications holds paramount importance. This objective can be realized through the formulation of models that are both computationally efficient and facile to deploy. Subsequent research endeavors should prioritize the simplification of these models' intricacy while upholding their efficacy. Potential strategies may encompass the utilization of approximation methodologies, streamlined mathematical representations, or innovative model structures customized for real-time computations.

Improving transfer learning frameworks to effectively manage inter-subject variability and decrease the requirement for substantial labeled data presents a significant challenge. Subsequent investigations ought to delve into resilient transfer learning methodologies capable of functioning efficiently with limited labeled data. Such endeavors may encompass the formulation of unsupervised or semi-supervised learning strategies, along with the utilization of domain adaptation techniques tailored for EEG data. Moreover, the exploration of more efficient approaches for aligning data across diverse subjects and sessions within the tangent space of the Riemannian manifold could contribute to enhancing the efficacy of transfer learning strategies.

To summarize, tackling computational intricacies, enhancing scalability, streamlining models for practical implementation, and refining transfer learning frameworks constitute pivotal domains for prospective studies. These endeavors are poised to facilitate the advancement of more effective, precise, and user-friendly Brain-Computer Interfaces (BCIs) that can find widespread utility in both clinical and non-clinical environments.

\section{Conclusion}
This study offers an extensive examination of the fusion of Riemannian geometry with deep learning methodologies within the domain of brain-computer interfaces. The unique amalgamation of these sophisticated mathematical strategies has illustrated substantial potential in augmenting the decoding capabilities of EEG signals in BCIs. Specifically, Riemannian geometry has played a pivotal role in the accurate and resilient management of the non-Euclidean data structures inherent in EEG data, an area where conventional Euclidean methods often prove inadequate. Our analysis has refreshed the landscape of EEG signal processing by presenting recent progressions that exploit deep learning to tackle and alleviate traditional BCI obstacles like sensitivity to noise, non-stationarity, and prolonged calibration periods. Through the perspective of Riemannian geometry, a shift has been observed in the utilization of covariance matrices, transitioning from fundamental manipulations to sophisticated metric learning and optimization for EEG classification. 

This transition not only enhances existing subspace techniques but also indicates a shift towards directly manipulating and categorizing covariance matrices, potentially circumventing traditional methodologies altogether. Nevertheless, despite the advancements, this analysis also highlights the necessity for further investigation to surmount the persisting constraints of BCIs. The infusion of Riemannian geometry into BCI design necessitates novel EEG representations that are more resilient and classifiers capable of effectively managing outliers and non-stationarity. Prospective avenues should also contemplate the formulation of adaptive techniques capable of dynamically adapting to the user's cognitive status and environmental variations. Additionally, delving into the possibilities of Riemannian geometry for multitask learning, feature extraction, and transfer learning within BCIs could yield substantial enhancements in both efficacy and usability. In summary, the convergence of Riemannian geometry and deep learning harbors significant potential for the trajectory of BCIs. 

As this domain progresses, it is imperative that sustained endeavors are channeled towards not only refining these mathematical frameworks but also rendering these sophisticated methodologies more accessible for pragmatic BCI implementations. We anticipate that the advancements expounded in this analysis will pave the way for more intuitive, efficient, and user-friendly BCI systems, setting a new benchmark for EEG signal classification and further bridging the chasm between theoretical exploration and real-world applications.

\end{document}